\documentclass[aps,prl,reprint,groupedaddress,showpacs,amsmath,amssymb,twocolumn,floatfix]{revtex4-1}
\usepackage{graphicx}
\usepackage{bm}
\usepackage{color}
\usepackage[normalem]{ulem}
\usepackage{cleveref}
\usepackage{amsmath}

\usepackage{amssymb}
\usepackage{epsfig}
\usepackage{epstopdf}
\usepackage{gensymb}
\usepackage[utf8x]{inputenc}
\usepackage{comment}

\begin{document}

\title{Magic continuum in twisted bilayer WSe$_{2}$}

\author{Lei Wang$^{1,2}$ $^{\dagger}$}
\author{En-Min Shih$^{2}$ $^{\dagger}$}
\author{Augusto Ghiotto$^{2}$ $^{\dagger}$}
\author{Lede Xian$^{3}$}
\author{Daniel A. Rhodes$^{1}$}
\author{Cheng Tan$^{1}$}
\author{Martin Claassen$^{4}$}
\author{Dante M. Kennes$^{3,5}$}
\author{Yusong Bai$^{6}$}
\author{Bumho Kim$^{1}$}
\author{Kenji Watanabe$^{7}$}
\author{Takashi Taniguchi$^{7}$}
\author{Xiaoyang Zhu$^{6}$}
\author{James Hone$^{1}$}
\author{Angel Rubio$^{3,4,8}$ $^{\ast}$}
\author{Abhay Pasupathy$^{2}$ $^{\ast}$}
\author{Cory R. Dean$^{2}$$^{\ast}$}

\affiliation{$^{1}$Department of Mechanical Engineering, Columbia University, New York, NY 10027, USA}
\affiliation{$^{2}$Department of Physics, Columbia University, New York, NY 10027, USA}
\affiliation{$^{3}$Max Planck Institute for the Structure and Dynamics of Matter, Luruper Chaussee 149,}
\affiliation{22761 Hamburg, Germany}
\affiliation{$^{4}$ Center for Computational Quantum Physics, Flatiron Institute, New York, NY 10010, USA}
\affiliation{$^{5}$ Institut f\"ur Theorie der Statistischen Physik, RWTH Aachen University, 52056 Aachen, Germany and JARA-Fundamentals of Future Information Technology, 52056 Aachen, Germany}
\affiliation{$^{6}$Department of Chemistry, Columbia University, New York, NY 10027, USA}
\affiliation{$^{7}$National Institute for Materials Science, Namiki 1-1, Tsukuba, Ibaraki 305-0044, Japan}
\affiliation{$^{8}$ Nano-Bio Spectroscopy Group, Departamento de Fisica de Materiales, Universidad del País Vasco, 20018 San Sebastian, Spain}
\affiliation{$^{\dagger}$These authors contributed equally to this work.}
\affiliation{$^{\ast}$Corresponding authors, Email: cdean@phys.columbia.edu, apn2108@columbia.edu, angel.rubio@mpsd.mpg.de}

\date{\today}



\maketitle


\textbf{
Emergent quantum phases driven by electronic interactions can manifest in materials with narrowly dispersing, i.e. ``flat'',  energy bands. Recently, flat bands have been realized in a variety of graphene-based heterostructures using the tuning parameters of twist angle, layer stacking and  pressure, and resulting in correlated insulator and superconducting states \cite{Cao2018, Cao2018a, Yankowitz2019, Sharpe605, lu2019superconductors, Serlin2019, Liu2019, Cao2019, Shen2019, Chen2019a, Chen2019, Serlin2019, chen2019tunable}. Here we report the experimental observation of similar correlated phenomena in twisted bilayer tungsten diselenide (tWSe$_{2}$), a semiconducting transition metal dichalcogenide (TMD). Unlike twisted bilayer graphene where the flat band appears only within a narrow range around a ``magic angle'', we observe correlated states over a continuum of angles, spanning 4$\degree$ to 
5.1$\degree$.  A Mott-like insulator appears at half band filling that can be sensitively tuned with displacement field.  
Hall measurements supported by \textit{ab initio} calculations suggest that the strength of the insulator is driven by the density of states at half filling,  consistent with a 2D Hubbard model\cite{Wu2018,Wu2019} in a regime of moderate interactions.  At 5.1$\degree$ twist, we observe evidence of superconductivity upon doping away from half filling, reaching zero resistivity around 3 K.  Our results establish twisted bilayer TMDs as a model system  to study interaction-driven phenomena in flat bands with dynamically tunable interactions.
}


The advent of van der Waals heterostructures \cite{Dean2010} has opened up new avenues to band engineering simply by placing one monolayer on top of another. When the two layers have different lattice constants, or the same lattice constant but are rotated with respect to each other, a long wavelength periodic modulation results, referred to as a moiré superlattice.  
Coupling to this moiré superlattice in general modifies the electronic bandstructure \cite{Park2008} and in some cases can result in the formation of new low-energy sub-bands\cite{Bistritzer12233}.
Under these circumstances, narrow and isolated sub-bands have been realized in a variety of graphene-based structures including twisted bilayer (TBG)\cite{Cao2018a, Cao2018, Yankowitz2019, Sharpe605, lu2019superconductors, Serlin2019}, double bilayer graphene(TDBG) \cite{Liu2019, Cao2019, Shen2019} and ABC trilayer graphene/BN(ATGB) \cite{Chen2019a, Chen2019, chen2019tunable}. Within these ``flat bands'' electron interactions become the dominant energy scale leading to new emergent electronic phases such as correlated insulators, superconductivity, magnetism and topological electronic structure. The question that naturally arises is how general these phenomena are to the whole class of van der Waals materials, and what kinds of new interaction-driven phases can be realized in other materials besides graphene.

\begin{figure*}[t]
\begin{center}
\includegraphics[width=0.9\linewidth]{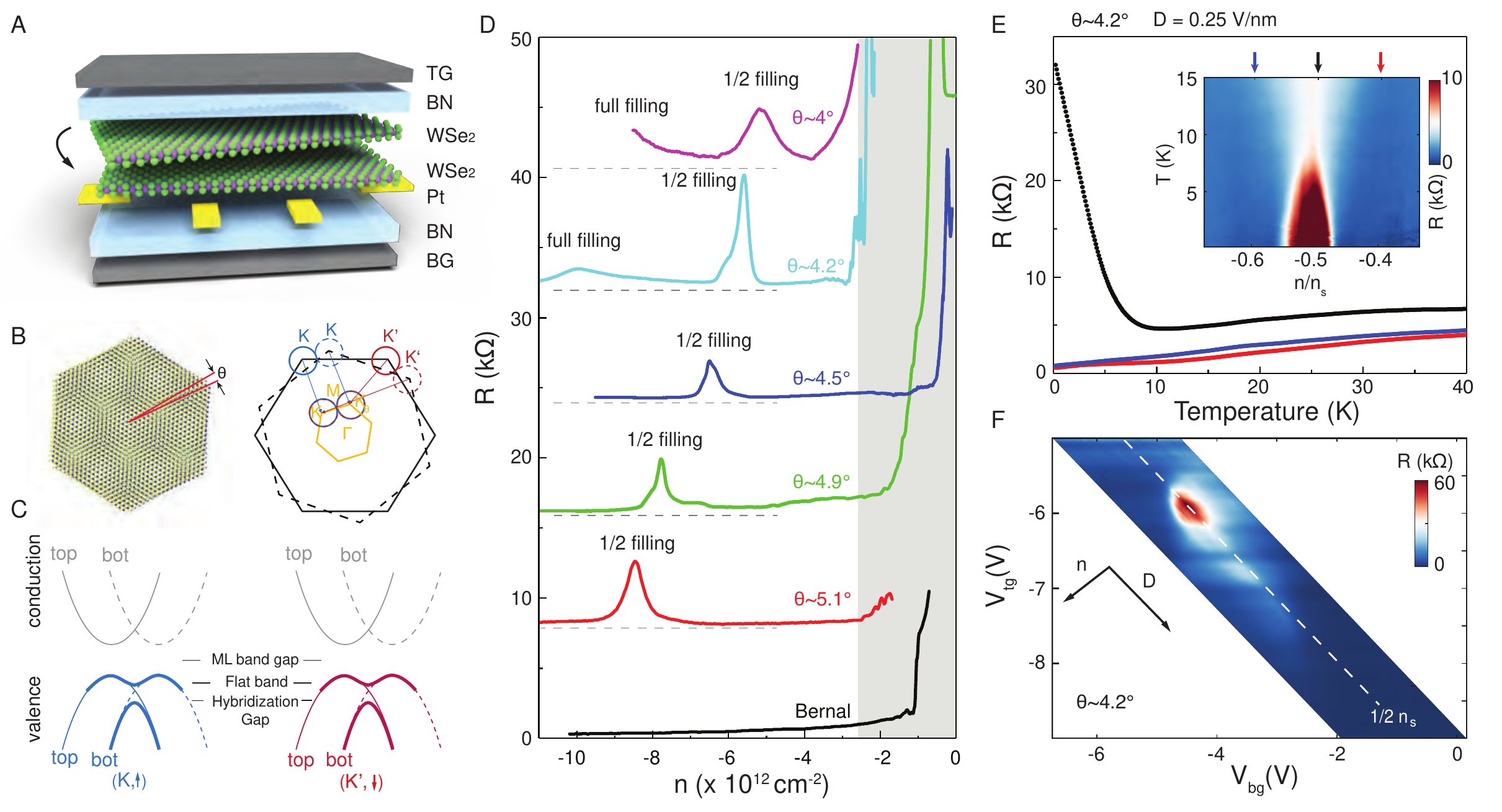}
\caption{\textbf{Flat bands in twisted bilayer WSe$_{2}$.} \textbf{A} Cartoon schematic of our device structure (see main text). \textbf{B} Left: Real space representation of the moiré pattern that results from a small angle twist between the two WSe$_2$ layers. Right: Brillouin zones of the top (solid black) and bottom (dashed black) layers. The states near K and K' are shown as blue and red, respectively. 
The resultant bandstructure in the moiré Brillouin zone in yellow is also displayed. \textbf{C} Schematic representation of the twisted WSe$_2$ hybridized bandstructure. Due to coupling of the spin and valley degeneracy, the resulting moiré valence band can host 2 electrons. \textbf{D} Resistivity curves as a function of total hole density for six WSe$_2$ bilayer configurations. ``Bernal" refers to the natural stacking arrangement (equivalent to a 60$\degree$ twist). 
\textbf{E} Resistivity versus temperature measured from the  4.2$\degree$ twist angle device highlighting the activation of the Mott gap and the metallic behaviour at densities away from 0.5n$_s$. Inset: resistivity map as a function of temperature and relative filling. \textbf{F} Evolution of the flat band response near half filling versus the top and bottom gate bias. Black arrows indicate the direction of increasing displacement field and carrier density.  The response at half filling shows a metal-insulator-metal transition under applied displacement field}
\label{fig:Fig1}
\end{center}
\end{figure*}

\begin{figure*}[t]
       \begin{center}
 \includegraphics[width=.9\linewidth]{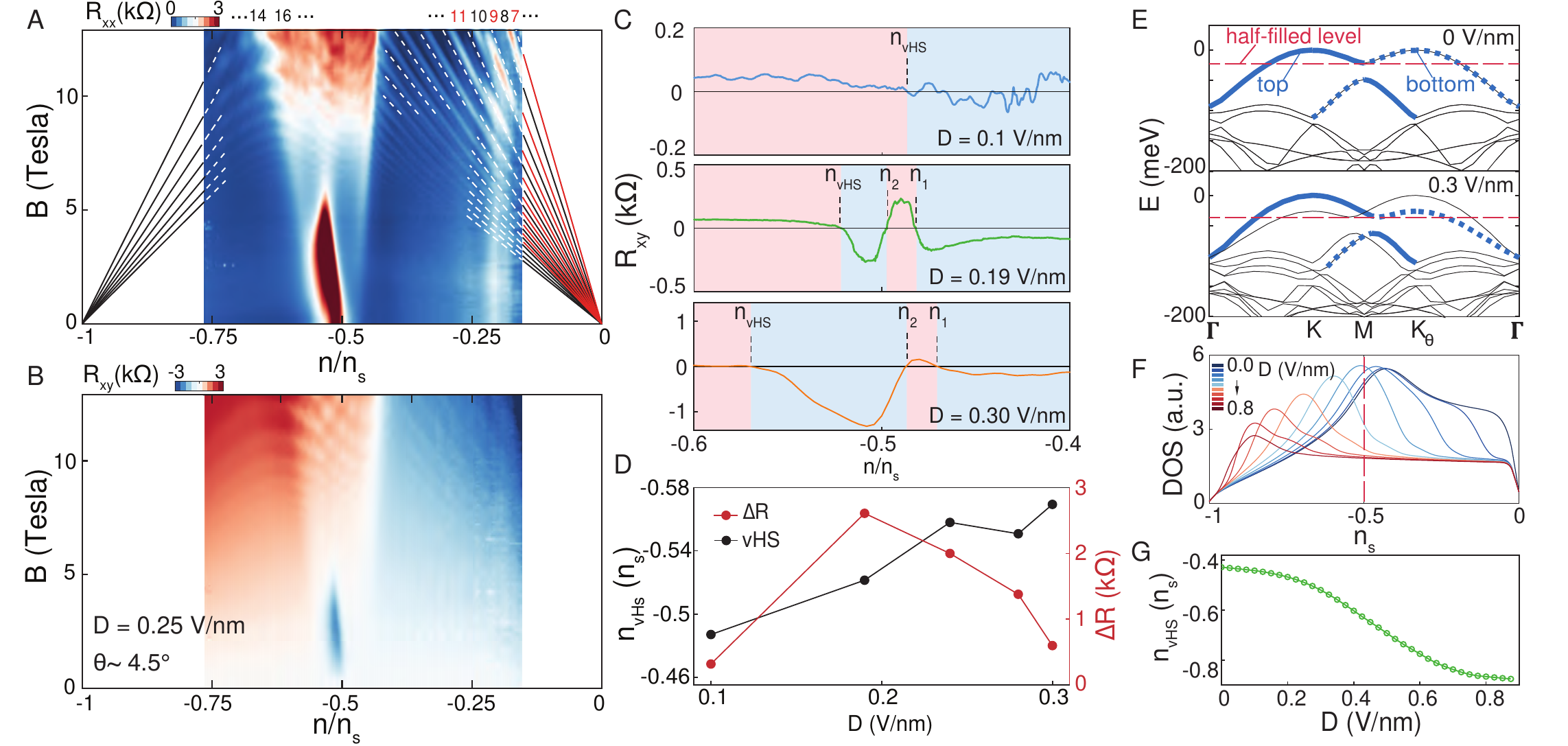}
		\caption{\textbf{Displacement field dependent van Hove singularity} \textbf{A}  Landau fans emerging from the band edge and full-filling of the moiré unit cell for a 4.5$\degree$ sample. Top gate is kept fixed at -19.5 V and the back gate tunes the density in the system as the magnetic field is swept and the longitudinal resistance is recorded. Dashed White and Red lines identify even and odd integer quantum Hall states, respectively.  Two landau fans are resolved, projecting (solid lines) to the empty and full filling densities of the moiré subband. 
		\textbf{B} Hall resistance versus field and filling. The Hall sign changes from hole-like to electron-like, consistent with doping of a single band (see text).  
		\textbf{C} R$_{xy}$ versus density, measured at $B=1$~T for various displacement fields (D). $n_{vHS}$ labels the position of the single-particle band van Hove singularity (vHS) for each $D$; $n_{1}$ and $n_{2}$ label additional sign changes that we associate with the correlated insulator (see text). \textbf{D} Left: Band filling position of the vHS versus $D$, determined from the Hall effect. Right: resistance of the half filling resistive peak versus $D$ for the same device ($\Delta R = R_{n/n_s = 0.5}-R_{n/n_s = 0.6}$). The peak resistance is maximal when the vHS is close to half filling  \textbf{E} DFT calculations of the top valence bands of the MBZ for $D=0$ and $D=0.3$~V/nm  fields for a 5.08$\degree$ twist. We highlight the contributions from the K valley of the top layer (solid blue) and from the K valley of the bottom layer (dashed blue) to the topmost bands. Red dashed lines indicate the energy level of half-filling of the topmost valence band. \textbf{F} Calculated density of states for the top two valence bands as a function of filling for a 4.5$\degree$ twist. 
		\textbf{G} Calculated position of the vHS position versus displacement field for a 4.5$\degree$ twist.
	}
		\label{fig:Fig2}
		\end{center}
\end{figure*}


\begin{figure*}[t]
	\begin{center}
	\includegraphics[width=0.9\linewidth]{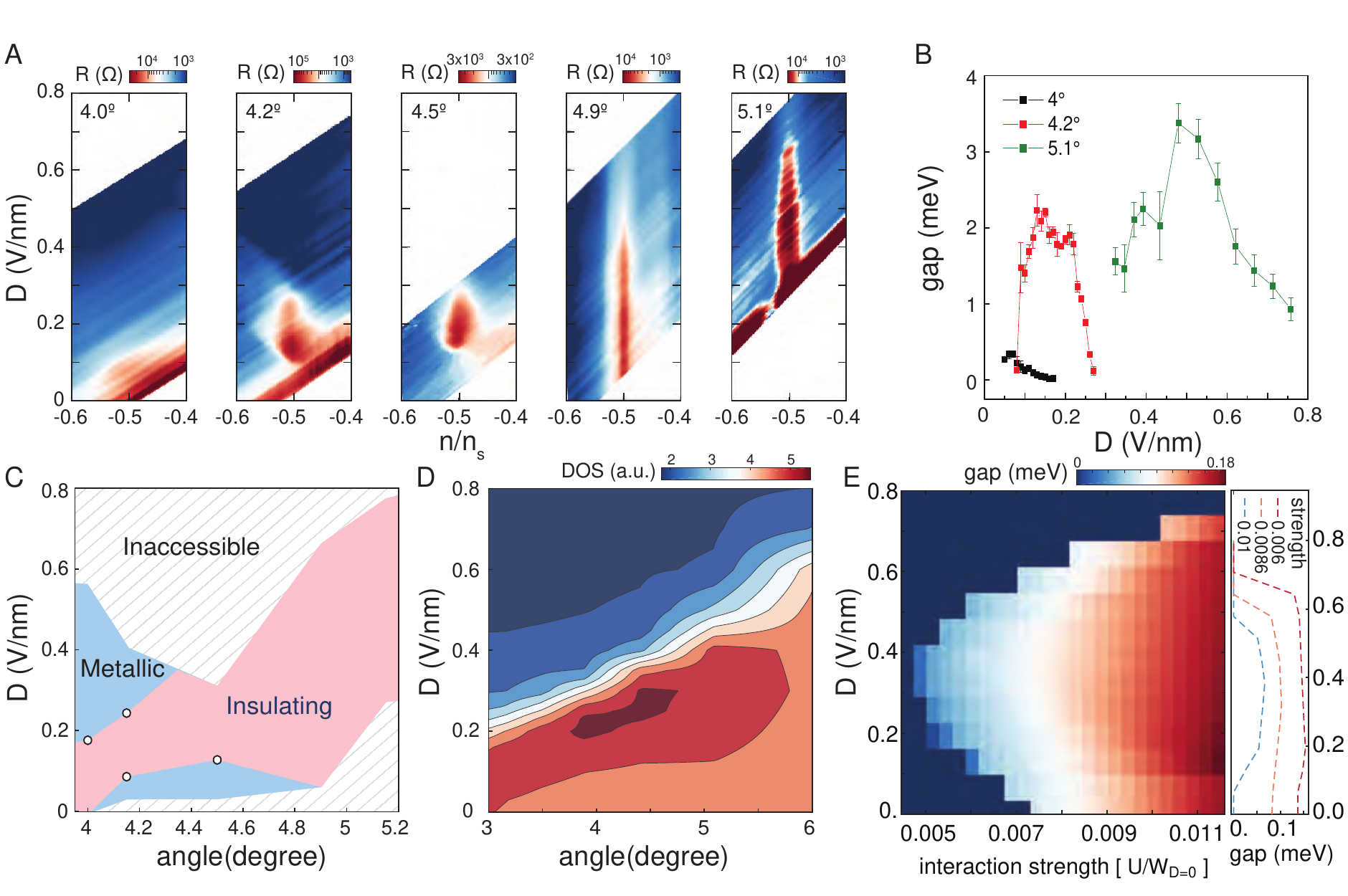}
		\caption{\textbf{Angle and displacement field dependence of the correlated insulator.} \textbf{A} Resistance versus displacement field and relative density near half-filling for all measured angles. \textbf{B} Summary of measured insulating gap sizes over displacement field across three devices. \textbf{C} Summary map of observed metallic and insulating regions at half-filling across different angles and displacement fields. White circles denote observed metal-to-insulator transitions. \textbf{D} Calculated density of states at half-filling for different angles and displacement fields, showing qualitative correspondence  with panel C. \textbf{E} Mean-field calculation of the gap size versus displacement field and assumed strength of the Coloumb interaction in a false color plot (left panel). The right panel shows line cuts for the gap size as a function of displacement field for a select interaction strengths. 
		}
		\label{fig:Fig3}
	\end{center}
\end{figure*}

\begin{figure}[ht]
	\begin{center}
	\includegraphics[width=0.9\linewidth]{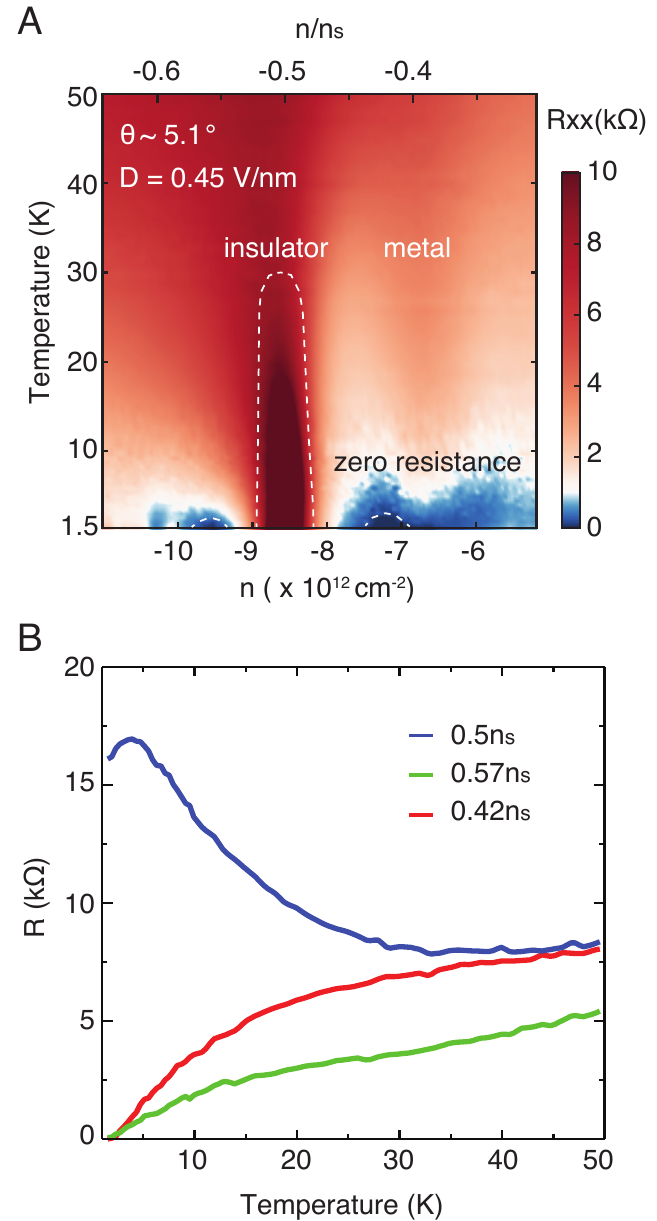}
		\caption{\textbf{Evidence of superconductivity} \textbf{A} Resistivity  versus temperature and density.  The density was varied with the back gate while the top gate was kept fixed at -12.25 V. Two zero resistance regions (suggesting superconducting behaviour) appear  when doping away from the insulating state. \textbf{B} Resistance versus temperature at different densities near half filling. The insulating response at half filling  (blue) onsets at ~35K, whereas the superconducting states (red and green) reach zero resistance below 3K. 
		}
		\label{fig:Fig3}
	\end{center}
\end{figure}

In this regard, the semiconducting 2H-transition metal dichalcogenides (TMDs) are a natural place to look for similar phenomena.  
Indeed, theoretical work on TMD homo- and hetero-bilayers predict the existence of flatbands and van Hove singularities (vHSs) in the moiré Brillouin zone (MBZ) \cite{Wu2018, Wu2019, Naik2018, Ruiz-Tijerina2019} that could support emergent electronic phases \cite{Wu2018, Wu2019, Schrade2019}. Moreover, the reduced degeneracy in TMDs compared to graphene, due to coupled spin and valley degrees of freedom, suggest that twisted bilayer TMDs could provide an idealized solid state version of a single band Hubbard model on a triangular lattice, where several exotic new states may be realized\cite{Wu2018,dagotto94,scalapino12,leblanc15,mazurenko17}.
In this work we report transport measurements of twisted bilayer WSe$_{2}$ for interlayer twist angles, $\theta$, spanning several degrees.  We observe a correlated insulator phase at half filling of the twist-induced mini-band,  as well as evidence of superconductivity in the vicinity of the insulating phase. Different from graphitic systems, these correlated states manifest over a wide range of twist angles \cite{Lede19}, while also showing strong tunability with displacement fields. 


Our devices were fabricated using  the same ``tear and stack'' method that has been used to fabricate twisted bilayer graphene with controlled interlayer twist angle \cite{Kim2016,Cao2018,Wang614}.  A cartoon illustration of the device structure is shown in Fig. 1a. In brief (see methods for details), our devices are assembled from high quality WSe$_2$ material, synthesized by the flux growth method\cite{Edelberg2019}, and employ pre-patterned Hall-bar shaped Pt electrodes \cite{Movva2015} together with a dual gate structure to achieve Ohmic contacts at low temperatures\cite{Movva2017}. The twist angles of measured devices were determined by a combination of techniques including atomic force microscopy, optically measured second harmonic generation and  transport signatures.  


 A schematic illustration of the moiré superlattice and resulting bandstructure for twisted bilayer WSe$_{2}$ is shown in Fig. 1b-c. In the vicinity of the K point of the Brillouin zone, the band structure of each WSe$_{2}$ layer is a single spin-polarized parabolic band. Electron hopping between the layers causes a strong hybridization of the bands where they intersect, leading to a low energy mini-band (Fig. 1c). 
For chemical potentials that are close to the valence band edge (as is the case in our samples), the physics of tWSe$_{2}$ is thus well approximated by a single subband whose width is determined by the twist angle. From angle-dependent bandstructure calculations, fitting over the range of 2$^{o}$ to 7$^{o}$ (Supplementary Information Fig.1), we find that the width of this subband, $w$ (in meV), grows approximately as $w\sim4.5\theta^2 - 5.4\theta + 7.1$ with twist angle $\theta$  in degrees. On the other hand, if we assume that the on-site Coulomb interaction $U$ between two electrons on the same moiré site is inversely proportional to the size of the moiré unit cell, then $U$  scales approximately linearly with $\theta$\cite{Cao2018}. This implies the existence of a crossover angle, below which $U$ exceeds $w$ and therefore electron correlations dominate.  We note however the precise angle that distinguishes strongly interacting from single particle regimes is  difficult to know \textit{a priori} due to ambiguities in the precise magnitude of the Coulomb integrals.


Fig.1d shows a plot of the resistivity versus hole-band carrier density at $T=1.8$~K for five different tWSe$_{2}$ devices with angles ranging from 4$\degree$ to 5.1$\degree$,  and also one naturally-occurring Bernal bilayer stacking corresponding to an equivalent twist of 60$\degree$ (curves are offset vertically for clarity).  All devices exhibit diverging resistance for hole densities less than $\sim2\times 10^{12} cm^{-2}$, due to the  Fermi energy approaching the intrinsic band edge of  WSe$_{2}$.  This sets a lower limit on the density range we can access by transport measurement.

At large densities we observe new resistive peaks, whose position shifts to lower density at smaller angles.  
The moiré unit cell area relates to the twist angle according to $Area=\frac{a^2\sqrt{3}}{4(1-cos\theta)}$, where $a= 0.328$nm is the WSe$_2$ lattice constant \cite{Kormanyos2015}. From this we observe that all devices exhibit a resistive peak that coincides with one hole per moiré unit cell, or half filling of the moiré subband (Supplementary Information Fig. 2). In addition, a second resistive peak is seen in the 4.2$\degree$ and 4$\degree$ devices near full fillng of the moiré subband (i.e. doping to 2 holes per unit cell).

 The appearance of the resistive features confirm the presence of a low energy band resulting from the twist-induced moiré superlattice\cite{Wu2018, Wu2019, Naik2018, Ruiz-Tijerina2019} (by comparison the Bernal device shows a featureless response with resistance monotonically decreasing  with increasing doping). The resistance peak near full filling is consistent with a reduction in the density of states (DOS) near the edge of the moiré subband\cite{Wu2018, Naik2018, Ruiz-Tijerina2019}. 
 In contrast, the  resistive state at half filling in the twisted devices is not anticipated in a single particle bandstrucure, and therefore indicates that correlations are sufficiently strong within this band to induce a spontaneously formed insulating state, similar to what has been observed previously at half filling of the flat bands in moiré-patterned graphitic systems\cite{Cao2018, Cao2018a, Yankowitz2019, Sharpe605, lu2019superconductors, Serlin2019, Liu2019, Cao2019, Shen2019, Chen2019a, Chen2019, Serlin2019, chen2019tunable}. The persistence of the correlated insulator (CI) state over more than a full degree of twist angles confirms that, unlike in twisted bilayer graphene, there is not a narrowly defined magic angle in which Coulomb energy dominates over the bandwidth but rather a broad continuum of angles.  

Fig.1e shows the temperature dependence near half filling for the 4.2$\degree$ device (Supplementary Information Fig. 3 for similar data from another device). We define the band filling as $\nu=n/n_{s}$ where $n$ is gate-induced hole density and $n_{s}$ is the density necessary to reach full filling of the moiré band.  Above approximately 10 K the sample shows a metallic response at all fillings  with resistance decreasing with decreasing temperature. However, below ~10~K the resistance at 0.5n$_{s}$ shows a sudden onset to thermally activated beahviour indicating an insulating reponse.  The extracted gap value is $\sim23$~K, or about twice the apparent critical temperature where the insulating response onsets.  The sharp transition together with an activation energy exceeding the onset temperature are both consistent with the CI being a  Mott transition\cite{Imada}.

Fig.1f shows the resistance of the same 4.2$\degree$ device versus top and back gate voltage (Supplementary Information Fig. 4 for other angle data) in the range of half filling, which allows us to investigate the effect of displacement field - here we report the average displacement field defined as $D=(C_{T}V_{T}-C_{B}V_{B})/2\epsilon_{o}$, where $V_{T(B)}$ is the top (bottom) gate bias, $C_{T(B)}$ the associated geometric capacitance, and $\epsilon_{o}$ is the vacuum permittivity.  We note that the resistive peak at 0.5$n_s$ persists only over a finite range 0.08~V/nm~$< D <$~0.28~V/nm, with the maximum resistance observed at $D \sim 0.15 V/nm$. Temperature dependent measurements (Supplementary Information Fig. 5) confirm a \textit{D}-field induced metal-insulator-metal transition over this range, with activation gaps showing a dome-like evolution.

The response of the samples under perpendicular applied magnetic fields provides key additional insight. We first map the Landau level spectra as a function of applied field and induced density to unambiguously determine the density of full filling of the moiré lattice. Next, we measure the Hall response to determine the point at which the band curvature changes from electron-like to hole-like, which should coincide with a peak (vHS) in the density of states. Fig. 2a shows the longitudinal resistance R$_{xx}$ for the 4.5$\degree$ device plotted against n/n$_{s}$ at magnetic fields up to 13 T. At high fields, two sets of Landau fans are visible. Linear extrapolation of the Landau level trajectories to zero field allows us to precisely determine the position of the valence band edge as well as the full filling density of the moiré subband (see Supplementary Information Fig. 6 for data from other devices and angles). 
For the fan originating from the valence band edge, we observe only even-integer quantum Hall states (white lines) up to approximately $B=8$~T, consistent with a two-fold degenerate Fermi surface.  Above 8~T, we observe onset of additional odd integer QHE states (red lines), indicating a lifting of the combined spin-valley degeneracy. This degeneracy lifting is not clearly observed in the satellite fan originating from full filling, where two-fold degenerate Landau levels persist to the highest measured fields. The magnitude of the 0.5$n_{s}$ insulating peak decreases with increasing B field and vanishes at B $\sim$ 6T. This behaviour is qualitatively consistent with a non-ferromagnetically ordered ground state. However, a detailed understanding of this trend is complicated by variation of both Coulomb and Zeeman energies under perpendicular applied field.  A full study of B-field dependence is beyond the scope of the present manuscript and will be discussed elsewhere (see also Supplementary Information Fig. 7).
Fig. 2b shows the Hall resistance  measured simultaneously with the longitudinal resistivity shown in Fig. 2a. At high field, the sign of the Hall resistance inverts upon doping from low to high carrier density, consistent with the expected response for a single  band where the dispersion changes from hole-like to electron-like as the band is filled. This provides further confirmation that our gate range spans a single low-energy sub-band. 
Fig. 2c shows the Hall resistance versus density, measured at $B=1$~T for three separate displacement fields . The top panel ($D=0.1$~V/nm) corresponds to low displacement field where no resistive peak is seen in the longitudinal resistance at half filling. Here, the Hall resistance changes sign from hole-like to electron-like transport at $n/n_s\sim0.4$, after which the Hall effect continues to show electron-like transport with increased doping. Since there is no evidence of an insulating sate, we interpret the single crossing point as the position of the van Hove singularity (vHS) where the moiré subband curvature changes sign. The two lower panels of Fig. 2c (D=0.19 V/nm and D=0.30 V/nm respectively), correspond to the range of $D$ where a resistive peak is observed at half filling. In this case three crossings appear, which we label $n_{1}$, $n_{2}$ and $n_{vHS}$.  We identify the $n_{vHS}$ crossing as the single-particle band vHS since this marks the point beyond which the band continues to show an electron-like Hall response. We therefore associate the other two crossings  with the appearance of the half-filling insulating state. 

The position of the vHS evolves continuously with displacement field, starting at $n/n_s<0.5$ for low $D$ but moving through half filling to  $n/n_s>0.5$ values for large $D$. The magnitude of the longitudinal resistance peak at half filling correlates closely with the position of the single-particle vHS with the largest resistive peak seen when the vHS is near half-filling (Fig. 2d). The dependence of the half filling CI on the vHS position indicates that the system is in a regime of moderate correlations, where the band structure (specifically the DOS) plays an important role in determining the properties of the emergent insulator. However, we note that the insulating phase itself is always observed at precisely half-filling, indicating that we are not in the weakly correlated regime where gaps would form at the peak of the vHS. 

 Fig. 2e-g shows the tWSe$_{2}$ band structure calculated by DFT as a function of $D$, for a 4.5$\degree$ twist. 
 At zero $D$, the two layers are degenerate and the hybridization of the valence bands results in a vHS exactly at the M point of the bandstructure. As a displacement field is applied to the sample, the bands from the two layers acquire differing onsite potentials. As a consequence, the position of the vHS, shifts away from the M point and down in energy (at extreme displacement fields, such as 1 V/nm, the two layers become effectively decoupled in the first moiré subband, but in practice, we do not reach such fields in experiment). Fig. 2f  shows the calculated DOS over the whole moiré subband. At $D=0$ the vHS (peak DOS) is located at slightly less than half filling. With increasing D, the vHS shifts thorough half filling and quickly moves towards full filling at large field. This trend is qualitatively consistent with our understanding of the Hall response under varying D (Fig. 2c-d), as discussed above. 
 Fig. 2g summarizes the evolution of the DOS at $n/n_s$=0.5 as a function of displacement field. The DOS at half filling varies non-monotonically with $D$,  initially increasing slightly and then dropping to about a third of the maximum value at larger fields (Supplementary Information Fig. 1). The correspondence between this trend and the non-monotonic evolution of the resistive peak (Fig. 1f), further supports our interpretation that the strength of the insulator strongly depends on the DOS.

Fig. 3a summarizes the behaviour of the correlated insulator (CI) for every twist-angle we measured. We observe a displacement field driven metal-to-insulator transition at half-filling for 4$\degree$, 4.2$\degree$ and 4.5$\degree$ twists, whereas for 4.9$\degree$ and 5.1$\degree$ the insulator-metal transition at low displacement field is not observed due to limitations on the achievable gate voltages. The evolution of the CI gap  versus displacement field, determined from thermal activation, is shown in fig. 3b.  For the 4.2$\degree$ and 5.1$\degree$ devices we confirm a non-monotonic gap evolution, peaking at a non-zero $D$ value. From these plots we construct a phase diagram for tWSe$_2$ as a function of angle and displacement field, shown in Fig. 3c. For comparison, Fig. 3d shows the DOS at half filling from single particle band structure calculations. A qualitative correlation is seen between the two plots, suggesting that the DOS at half filling plays an important role in stabalizing  over the entire angle range we measure. 

In addition to DFT we performed mean field calculations of the correlation gap (see methods for details).
 In the limit of weak interactions, the calculated CI gap shows a non-monotonic dome-like $D$-field dependence, with a peak  value occurring where the vHS lies on the Fermi surface (Fig. 3e). This is qualitatively in agreement with the correspondence between Figs. 3c and 3d, as discussed above. For larger interactions, the gap is nearly constant  from zero $D$ up to a critical value, beyond which the system becomes metallic. The displacement field dependence seen in experiment is best modelled by choosing a Hubbard interaction $\sim 0.5-0.7$meV. This would place the system firmly in the weak coupling regime $U\ll W$.  
 However this is inconsistent with the insulator being pinned to half-filling case as seen in experiment, and may be a consequence of the quantitative overestimation of the ordering tendency via mean-field treatment. To arrive at a quantitatively predictive solution, future theoretical work should address the computationally challenging intermediate coupling regime as well as characterize the role of a displacement field dependent interaction strength from \textit{ab initio} simulations.

Finally we investigate whether correlations in doped tWSe$_{2}$ can lead to the emergence of superconductivity\cite{Cao2018a, Yankowitz2019, lu2019superconductors, Liu2019, Shen2019, Chen2019a}.  Fig. 4 shows the temperature dependence of the the 5.1$\degree$ sample in the vicinity of half filling. This data was acquired at D=0.45 V/nm - corresponding to the condition of strongest CI observed for any of the devices measured (see Fig. 3b).  The resistance was measured using the smallest currents we could source in practice (5 nA) at this displacement field. On either side of the CI state, we observe regions in which the resistance reaches zero within our instrumentation noise of $\sim10$~$\ohm$, suggesting a superconducting transition. Fig. 5b shows the resistivity versus $T$ at half filling, where an insulator behaviour is seen at low $T$, as well as at two densities where the zero-resistance state is observed. For all three densities, the temperature dependence shows a linear-in $T$ metallic behavior above $\sim$40~K.  At half filling, thermally activated behaviour onsets below 35 K, whereas at the densities where superconductivity is observed, the resistance begins to deviate from its high temperature linear slope at around 15 K, dropping finally to zero at around 3 K (see also Supplementary Information Fig. 8). The true zero resistance state appeared to be fragile and was unstable to repeated thermal cycling, possibly indicating significant inhomogeneity in the superconducting phase. For other angles, we also observe nonlinear drops in resistance in the vicinity of half filling, suggestive of incipient superconductivity, but the minimum resistance in these samples at 1.5 K did not reach below $\sim200 \ohm$  (see Supplementary Information Fig. 3). 

In conclusion, we demonstrate the ability to engineer flat bands in twisted bilayer WSe$_{2}$ over a continuous range of twist angles.  We show that this system hosts correlation-driven emergent phenomena similar to graphitic moiré systems, but with very different degeneracies and single-particle bandstructures. Due to their easily tunable  electronic structure, this provides a model system for exploring other emergent electronic phases such as exciton condensates, spin liquids, and magnetic ordering.  More generally the system provides a tunable solid state realization of a single band Hubbard model on a triangular system where bandwidth and doping can be be independently varied, allowing new opportunities for studying the interplay between strong interactions and frustration. 

\bibliographystyle{apsrev4-1}
\bibliography{refs.bib}

\bigskip
\section*{Acknowledgments}
Studies of the tunable correlated states in twisted bilayer WSe$_{2}$ were supported as part of Programmable Quantum Materials, an Energy Frontier Research Center funded by the U.S. Department of Energy (DOE), Office of Science, Basic Energy Sciences (BES), under award DE-SC0019443.  Synthesis of WSe$_{2}$ crystals was supported by the NSF MRSEC program through Columbia in the Center for Precision Assembly of Superstratic and Superatomic Solids (DMR-1420634). Theoretical work was supported by the European Research Council (ERC-2015-AdG694097), cluster of Excellence AIM, SFB925 and Grupos Consolidados (IT1249-19). The Flatiron Institute is a division of the Simons Foundation.  We acknowledge support from the Max Planck—New York City Center for Non-Equilibrium Quantum Phenomena.

\bigskip
\section{Author Contributions}
L.W., E.S. and A.G. contributed equally to this work. L.W., E.S., A.G., C.T. and D.R. fabricated the samples. L.W., E.S. and A.G. performed the transport measurements and analyzed the data. Y.B. and X.Y.Z. performed SHG measurements. D.R., B.K. and J.H. grew the WSe$_2$ crystals. K.W. and T.T. grew hBN crystals. L. X. performed DFT and tight-binding calculations. M.C. and D.M.K. did mean-field calculations. A.R. supervised the theoretical aspects of this work. C.R.D, A.P., L.W., E.S. and A.G. wrote the manuscript with input from all authors.

\bigskip
\section{Competing interest declaration}
The authors declare no competing interest.

\newpage


\end{document}